\documentclass[aps, showpacs]{revtex4}
\usepackage{amsfonts,amsmath,amssymb,bm,graphicx}

\begin{document}

\title { Conformal symmetry algebra of the quark potential
and degeneracies in the  hadron spectra  }

\author{M. Kirchbach}
\affiliation{Instituto de F\'{\i}sica, Universidad Aut\'onoma de San Luis Potos\'{\i},
Av. Manuel Nava 6, San Luis Potos\'{\i}, S.L.P. 78290, M\'exico}

\begin{abstract}
The essence of the potential algebra concept {\it [Y.\ Alhassid, F.\  G\"ursey, F.\  Yachello. Phys.\ Rev. \ Lett. {\bf 50} (1983)]}
is that quantum mechanical free motions of scalar particles on curved surfaces of given
isometry algebras can be mapped on 1D Schr\"odinger equations
with particular potentials. As long as the Laplace-Beltrami operator on a curved surface  is
proportional to one of the Casimir invariants of the isometry algebra, free motion on the surface
is described by means of the eigenvalue problem of that very Casimir operator. In effect, 
the excitation modes considered are classified according to
the irreducible representations of the algebra of interest and are characterized by typical degeneracies.
In consequence, also the spectra of the equivalent  Schr\"odinger operators  
are classified according to the same irreducible representations and carry the same typical degeneracies. 
A subtle point concerns the representation of the algebra elements which may or may not be unitarily equivalent to the
standard one generating classical groups like  $SO(n)$, $SO(p,q)$, etc. 
To be specific, any similarity transformations of an algebra that underlies, say, an orthogonal group, 
always conserve the commutators among the elements, but a non-unitarily transformed algebra must not generate same group. 
One can then consider the parameters of the non-unitary similarity transformation as group symmetry breaking scales and seek to identify them
with physical observables.
We here use the potential algebra concept as a guidance in the search for an
interaction describing  conformal degeneracies. For this purpose we subject the $so(4)\subset so(2,4)$ isometry
algebra of the $S^3$ ball to a particular non-unitary similarity transformation and
obtain a deformed isometry copy to $S^3$ such that free motion on the copy is equivalent to a cotangent perturbed
motion on $S^3$, and to the 1D Schr\"odinger operator with the trigonometric Rosen-Morse potential as well. 
The latter presents itself especially well suited for quark-system studies insofar as its Taylor series decomposition begins
with a Cornell-type potential and in accord with lattice QCD predictions.   
We  fit the  strength of the cotangent potential  to the spectra of
the unflavored high-lying mesons and obtain a value compatible with the light dilaton mass. 
We conclude that while the conformal group symmetry of QCD following from $AdS_5/CFT_4$ may be broken by the dilaton mass, 
it still may be preserved as a symmetry algebra of the potential, thus explaining the
observed conformal degeneracies in the unflavored hadron spectra, both baryons and mesons.
\end{abstract}

\pacs{ 12.39.Jh, 24.85.+p}
\keywords{conformal potential algebra, $AdS_5/CFT_4$  duality, dilaton mass, super-symmetric degeneracies, 
unflavored  hadron spectra }

\maketitle

\begin{flushright}
{\it Explorer,  your  footsteps  are  the trail,\, and \, nothing else: }\\
{\it explorer,\,no\, trail \,to\, follow, advancing  trails  are blazed.\, }\\
{\it after Antonio Machado\,}
\end{flushright}

\begin{flushright}
{\it To the 70th birthday of Francesco Yachello}
\end{flushright}

\section{The potential algebra concept}
Group theoretical approaches to both  bound and scattering state problems in physics have  
played a pivotal  r\'ole in our understanding of
spectra classifications. It is a well known fact, that several of
the exactly solvable quantum mechanical potentials give rise to spectra which
fall into the irreducible representations of certain  Lie group algebras.
As a representative case, we wish to mention the widely studied class of 
Natanzon potentials \cite{Natanzon} known to produce spectra that 
populate  multiplets of the pseudo-rotational algebras so(2,2)/ so(2,1) \cite{Gango}. 
This phenomenon has been well understood in casting the Hamiltonians 
with the potentials under consideration as Casimir invariants of so(2,2)/so(2,1) algebras in representations 
not necessarily unitarily equivalent to the pseudo-rotational and 
referred to as ``potential algebras'' \cite{PotAl}- \cite{Quesne}.
The hyperbolic P\"oschl-Teller--, Scarf--,  and Eckart potentials stand for three popular interactions  having $su(1,1)$ symmetry algebras.

\subsection{The so(2,1) isometry algebra of the one-sheeted hyperbolic plane and the 
$\frac{-\lambda(\lambda +1)}{\cosh^2\rho}$  potential}
The essence of the potential algebra concept pioneered in \cite{PotAl}  is that quantum mechanical free motions 
on curved surfaces of given
isometry algebras can be transformed, upon an appropriate change of variables, into 1D  Schr\"odinger equations
with particular potentials. As long as the Laplace-Beltrami operator on a curved surface  is
proportional to one of the Casimir invariants of the associated isometry algebra, free motion on the surface
is described by means of the eigenvalue problem of that very Casimir invariant. In consequence,
the spectrum of the related potentials will be  classified according to the irreducible representations of the algebras 
under consideration and will be characterized by typical degeneracies. 
In particular, in  \cite{PotAl} attention has been drawn to the fact that the bound and scattering states of the
hyperbolic  $V=-\lambda(\lambda +1)/\cosh^2\rho $  potential (it can be part of either the  P\"oschl-Teller--, or the Scarf interactions),
fall into irreducible  representations of the $su(1,1)\sim so(2,1)$ algebra because
this potential can be embedded into the Casimir invariant, ${\mathcal C}$, of the pseudo-rotational algebra, 
$su(1,1)\sim so(2,1)$, acting as an isometry algebra of the one-sheeted  hyperboloid, $x^2+y^2-z^2=1^2$, an $AdS_2$ space. The algebra is
spanned by the generators
\begin{eqnarray}
\left[ J_z,J_\pm\right]=\pm J_\pm, &\quad& \left[ J_+, J_-\right]=-2J_z,\quad {\mathcal C}=J_z^2+J_z-J_-J_+=-\Delta,
\label{PT1}
\end{eqnarray}   
with $\Delta$ being the Laplacian on the one sheeted-hyperboloid \cite{Kalnins}.
Upon a parametrization of the above surface as 
\begin{eqnarray}
x=\cosh\rho \cos\varphi, &\quad& y=\cosh \rho \sin\varphi, \quad z=\sinh \rho,\quad
{\mathcal C} =\frac{1}{\cosh \rho }\frac{\partial }{\partial \rho }\cosh\rho \frac{\partial }{\partial \rho } +
\frac{m^2}{\cosh^2 \rho}, 
\end{eqnarray}
the quantum mechanical free motion of a scalar particle of mass $\mu$ 
on it is described by the $\left( -\frac{\hbar^2}{2\mu}\right){\mathcal C}$ eigenvalue problem as,
\begin{eqnarray}
-\frac{\hbar^2}{2\mu }{\mathcal C}{\mathcal Z}_k^m( \rho, \varphi)&=&-\frac{\hbar^2}{2\mu}k(k-1){\mathcal Z}_k^m (\rho ,\varphi), 
\quad \frac{\partial}{\partial \varphi}{\mathcal Z}_k^m(\rho, \varphi) =m  {\mathcal Z}_k^m(\rho, \varphi),
\label{PT2}
\end{eqnarray}
where ${\mathcal Z}_k^m(\rho, \varphi)$ are the $so(2,1)$ representation functions, while $k$, and $m$ stand for the corresponding representation
labels.
In the following, we chose to work in the dimensionless units, $\hbar=1$, $2\mu =1$.
In changing now variables to ${\mathcal Z}_k^m(\rho, \varphi) =\frac{U_k^m(\rho, \varphi)}{\sqrt{\cosh\rho}}$, converts  (\ref{PT2}) to the $1D$ 
Schr\"odinger equation with the  $\frac{-\lambda (\lambda +1)}{\cosh^2\rho}$ potential,
\begin{eqnarray}
{\mathcal H}U_k^m(\rho, \varphi)  = \epsilon_k U_k^m(\rho, \varphi), &\quad&
{\mathcal H} =-\frac{\partial^2}{\partial\rho^2}  -\frac{m^2-\frac{1}{4} }{\cosh^2\rho }+\frac{1}{4}, \quad \lambda=|m|-\frac{1}{2},\nonumber\\
\epsilon_k=-(|m|-\frac{1}{2}-n)^2+\frac{1}{4} &=&-k(k-1), \quad   k=|m|-n.
\label{PT3}
\end{eqnarray}
The $k$ label can be fixed to a positive integer which would mean choosing an infinite unitary discrete series representation of $so(1,2)$
with $|m|=k+n$, and $n=0,1,2,..$. Stated differently, the states bound within hyperbolic potentials of the type
$\left[-(|m|+1/2)(|m|-1/2)/\cosh^2\eta\right] $ with $|m|=k,k+1,...$ will populate  discrete unitary irreducible representations of $so(2,1)$.  
This is explained by the circumstance  that the associated Schr\"odinger Hamiltonian allows for a map on a linear function of the
$su(1,1)\sim so(2,1)$ Casimir operator according to,
\begin{eqnarray}
\frac{1}{\sqrt{\cosh\rho}}{\mathcal H}U_k^m(\rho, \varphi)&=&-\left( {\mathcal C} +\frac{1}{4}\right)\frac{U_k^m(\rho, \varphi)}{\sqrt{\cosh\rho}}
=-\left[k(k-1)+\frac{1}{4} \right]\frac{U_k^m(\rho, \varphi)}{\sqrt{\cosh\rho}}. 
\end{eqnarray}
One says, that the hyperbolic $\frac{-\lambda(\lambda+1)}{\cosh^2\rho}$ potential
has $su(1,1)\sim so(2,1)$ symmetry algebra, or, equivalently, that the isometry algebra $su(1,1)\sim so(2,1)$ 
of the one-sheeted hyperboloid is the potential algebra of the $\frac{-\lambda (\lambda +1)}{\cosh^2\rho}$ interaction
for the particular $\lambda =\left( |m|-\frac{1}{2}\right)$ values. The scattering states would be classified according to continuous-series representations of 
$su(1,1)$.
\subsection{The so(1,2) isometry algebra of the two-sheeted hyperbolic plane and the Eckart potential}
In a  similar way, the hyperbolic 
$\frac{\lambda(\lambda -1)}{\sinh^2\eta}$ potential can be related to  the Casimir invariant of the isometry algebra
of the two-sheeted hyperboloid, $z^2-x^2-y^2=1$, which is  $so(1,2)$, and can be parametrized as \cite{Kalnins}
\begin{eqnarray}
x=\sinh\eta \cos\varphi, &\quad& y=\sinh\eta   \sin\varphi, \quad z=\cosh \eta,\quad
{\mathcal C} =\frac{1}{\sinh \eta }\frac{\partial }{\partial \eta }\sinh \eta \frac{\partial }{\partial \eta } -
\frac{m^2}{\sinh^2 \eta}.
\label{ET}
\end{eqnarray}
Also in this case, the quantum mechanical free motion is described
in terms of the  eigenvalue problem of the relevant Casimir operator,${\mathcal C}$,
\begin{eqnarray}
-\frac{\hbar^2}{2\mu }{\mathcal C}Y_k^{m}(\eta, \varphi)&=&-\frac{\hbar^2}{2\mu}k(k+1)Y_k^{m} (\eta ,\varphi), 
\quad Y_k^{m}(\eta, \varphi)=P_k^{|m|}(\cosh \eta)e^{im\varphi},
\label{ET1}
\end{eqnarray}
where $Y_k^m(\eta, \varphi)$ are the pseudo-spherical harmonics. Also here, a suited variable change,
\begin{eqnarray}
Y_k^m(\eta, \varphi)=\frac{U_k^m(\eta, \varphi)}{\sqrt{\sinh\eta}}, 
\label{ET2}
\end{eqnarray}
converts the free motion on the upper sheet, ${\mathbf H}_+^2$, of the hyperbolic plane, into an 1D Schr\"odinger equation, this time with
the  Eckart potential \cite{Manning_Rosen} according to
\begin{eqnarray}
{\mathcal H}U_k^m(\eta, \varphi)  = \epsilon_k U_k^m(\eta, \varphi), \quad
{\mathcal H} &=&-\frac{\partial^2}{\partial\eta^2}  +\frac{m^2-\frac{1}{4} }{\sinh^2\eta }+\frac{1}{4},
\quad \lambda=|m|+\frac{1}{2},\nonumber\\
\epsilon_k=-(|m|+\frac{1}{2}+n)^2+\frac{1}{4} =-k(k+1), \quad   k=|m|+n,&\quad&
n,k=0,1,2,..., \quad |m|\in [0,k].
\label{ET3}
\end{eqnarray}
Also in this case, the conclusion can be drawn that the isometry algebra of the surface on which the free motion takes place, acts
as a symmetry algebra of the potential appearing in the equivalent Schr\"odinger equation.
However, a curious and most interesting situation occurs upon perturbing the aforementioned free motion by a $\coth\eta$ interaction, which results in 
\cite{Gango}
\begin{eqnarray}
-\frac{\hbar^2}{2M}\left( {\mathcal C}+2b \coth\eta \right) F_t^{\widetilde m}(\eta,\varphi)
&=&-\frac{\hbar^2}{2M}\epsilon_t F_t^{\widetilde m}(\eta,\varphi),\quad
\epsilon_t=t(t+1) +\frac{b^2}{\left(t+\frac{1}{2}\right)^2}, \quad t=n+|{\widetilde m}|.
\label{Wu5}
\end{eqnarray}
Despite non-commutativity of the perturbance with ${\mathcal C}$, the wave functions, $F_t^{\widetilde m}(\eta,\varphi)$,  in (\ref{Wu5}) continue 
transforming according to the finite-dimensional non-unitary representations of the $so(1,2)$ algebra,
and the corresponding excitations  are characterized by same degeneracies, as the kinetic motion. This becomes apparent upon changing
variables in (\ref{Wu5}) according to (\ref{ET2}), in which case one arrives at the well known
1D Schr\"odinger equation with the Eckart potential. 
In \cite{Nemy} this peculiarity of the Eckart potential has been explained by showing that
the Hamiltonians of the $\coth\eta$-perturbed--, and the free motions on ${\mathbf H}_+^2$ are related 
by a non-unitary similarity transformation, that can be directly read off from the decomposition
of the perturbed  solutions in the basis of the free ones. In dimensionless units one finds,
\begin{eqnarray}
({\mathcal C} +2b\coth \eta) {\mathbf  X}_t(\eta ,\varphi)&=&
\left[ {\widetilde {\mathcal C}} +\frac{\alpha^2_t}{4}\right]
{\mathbf  X}_t(\eta ,\varphi),\quad
{\widetilde {\mathcal C}}=
e^{-\frac{\alpha_t\eta }{2}}A_t (\varphi) {\mathcal  C} 
e^{\frac{\alpha_t\eta }{2}}A_t^{-1} (\varphi),\quad \alpha_t=\frac{2b}{t+\frac{1}{2}},\nonumber\\
{\mathbf X}_t(\eta, \varphi)&=&
e^{-\frac{\alpha_t \eta }{2}}\left( \begin{array}{c}
F_t^0(\eta)e^{i0\varphi}\\
F_t^1(\eta)e^{i\varphi }\\
...\\
F_t^t(\eta)e^{il\varphi}
\end{array}
\right)=
A_t(\varphi)
e^{-\frac{\alpha_t \eta }{2}}
\left( \begin{array}{c}
{ Y}_t^0(\eta, \varphi )\\
{  Y}_t^1( \eta, \varphi )\\
...\\
{  Y}_t^t( \eta ,\varphi )
\end{array}\right).
\label{bing_bong}
\end{eqnarray}
Here, $A_t(\varphi)$ are matrices,
\begin{eqnarray}
A_t(\varphi)&=&\left(
\begin{array}{cccc}
a_{11}^t&a_{12}^te^{-i\varphi} &...&a_{1(t+1)}^te^{-it\varphi}\\
0 &a_{22}^t&...&a_{2(t+1)}^te^{-i(t-1)\varphi }\\
...&...&...&...\\
0 &0 & 0&a_{(t+1)(t+1)}^t
\end{array}
\right),
\label{expansions}
\end{eqnarray}
which determine  the decomposition of the representation functions, ${\mathbf X}_t(\eta,\varphi)$,
of the similarity transformed Casimir operator, ${\widetilde {\mathcal C}}$, 
into the basis of the pseudo-spherical harmonics, the representation functions of the algebra underlying the pseudo-orthogonal 
group $SO(1,2)$. As long as the $t$ label takes same values as $k$ in (\ref{ET3}), and $|{\widetilde m}|$ obeys same branching 
rule with respect to $t$ as $|m|$ to $k$, the $(2t+1)$-fold degeneracies of the full Eckart potential are indistinguishable 
from the $(2k+1)$-fold ones of the free motion on ${\mathbf H}_+^2$. Important, the level splittings will be modified through the presence of
the representation constant $\alpha_t$ in (\ref{bing_bong}). Such type of differences
between pseudo-rotational spectra reveal changes in the representation of the algebra.
Explicit expressions for the  $A_t(\varphi )$  matrices can be found in \cite{Nemy}.
It is also important to be aware of the fact  that eq.~(\ref{bing_bong}) is not valid at the operator level, i.e.
$\left({\mathcal C} +2b\coth \eta \right)\not=\left( {\widetilde {\mathcal C}} +\frac{\alpha^2_t}{4}\right)$.
Its validity at the level of the eigenvalue problems is guaranteed by virtue of certain class of
recurrence relations among associated Legendre functions, listed in \cite{Nemy}.  
The non-unitary similarity transformation in (\ref{bing_bong}), (\ref{expansions}) is distinct from
the one considered in \cite{WuAlhassid} whose main purpose has been to prove that the $\coth\eta$-perturbed motion
on ${\mathbf H}_+^2$ gives rise to a Natanzon class potential.
Notice that the $\coth\eta $-perturbed motion on ${\mathbf H}_+^2$ is equivalent to free motion on the
$\exp (-2b\eta )$ rescaled hyperbolic space, the lowest solution of (\ref{Wu5}),  visualized in  Fig.~1. 
\begin{figure}
  \includegraphics[height=.1\textheight]{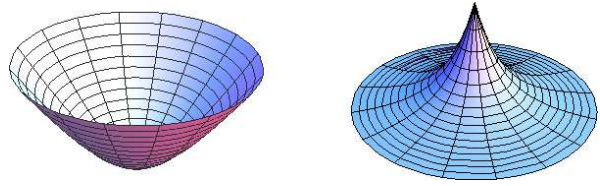}
 \caption{The $\coth\eta $ perturbed motion on ${\mathbf H}_+^2$ (left) is equivalent to free motion on the 
deformed surface (right).Obviously, the figure on the right is not a symmetric space of the $SO(1,2)$ group.
We conclude that though the Eckart potential has $so(1,2)$ as a symmetry algebra, the $SO(1,2)$ group symmetry of
the free motion has been broken by the scale introduced by the potential strength, causing a deformation of the ${\mathbf H}_+^2$
metric.
 }
\end{figure}\subsection{The so(3) isometry algebra of $S^2$ and the cotangent perturbed rigid rotator}
As long as $so(1,2)$, and $so(3)$ are related  by a  Wigner  rotation, 
${\mathcal C}$ in eq.~(\ref{ET}) and standard squared orbital angular momentum ${\mathbf L}^2$,
\begin{equation}
{\mathbf L}^2=
-\frac{1}{\sin\theta }\frac{\partial }{\partial \theta }\sin \theta
\frac{\partial }{\partial \theta } -\frac{1}{\sin^2\theta }
\frac{\partial^2}{\partial\varphi^2},
\label{oam}
\end{equation}
are related  by a complexification of the polar angle,
\begin{equation}
{\mathbf L}^2\stackrel{\theta \to i\eta }{\longrightarrow}{\mathcal C}.
\label{cmplxfctn}
\end{equation}
It is the type of complexification that takes  the sphere $S^2$ to the {\bf H}$^2_+$ hyperboloid .
Correspondingly, the properly altered  scaling transformation in eq.~(\ref{bing_bong}), together with $b\to ib$,
will take the free  motion on  $S^2$ to the one perturbed by  $\cot\theta$, 
the trigonometric counterpart of $\coth\eta$.
The $\cot\theta $ potential on $S^2$ is exactly solvable \cite{MolPhys} according to,  
\begin{eqnarray}
\left[ {\mathbf L}^2-
2b\cot \theta \right]{\mathcal X}_t^{\widetilde{m}}(\theta, \varphi) &=& 
\left[ \widetilde{\mathbf L^2}  -\frac{\alpha_t^2}{4}\right]{\mathcal X}_t^{\widetilde{m}}(\theta, \varphi)=
\left[ t(t+1) -\frac{\alpha_t^2}{4}\right] {\mathcal X}_t^{\widetilde{m}}(\theta, \varphi),\nonumber\\
\widetilde{{\mathbf L}^2}&=&
e^{-\frac{\alpha_t\theta }{2}}B_t (\varphi) {\mathcal  C} 
e^{\frac{\alpha_t\theta }{2}}B_t^{-1} (\varphi),\quad \alpha_t=\frac{2b}{t+\frac{1}{2}}.
\label{MPH1}
\end{eqnarray}
The $B_t(\varphi )$ matrices are the trigonometric analogues to $A_t(\varphi)$ in (\ref{bing_bong}) and (\ref{expansions}).
Also in this case, the wave functions of the perturbed motion continue transforming according to
$so(3)$ irreps but now represented by ${\mathcal X}_t^{\widetilde{m}}(\theta, \varphi)$, the eigenfunctions  
of the Casimir invariant, ${\widetilde{\mathbf L}^2}$, of the non-unitarily transformed $so(3)$ algebra,
\begin{eqnarray}
{\mathcal X}_t^{\widetilde{m}}(\theta, \varphi) ={\mathcal F}_t^{|\widetilde {m}|}(\theta )e^{i{\widetilde m}\varphi}
&\quad&  t=0,1,2,.. , \quad |{\widetilde m}|\in [0,t],
\label{MPH2}
\end{eqnarray}
where
\begin{eqnarray}
{\mathcal F}_t^{|{\widetilde m}|}(\theta ) &=&N_{t|{\widetilde m}|}e^{-\frac{b\theta }{t+\frac{1}{2}}} \sin ^{t}\theta
R_{n=t-|\widetilde{m}|}^{  \frac{2b}{t+\frac{1}{2}}, -(t-\frac{1}{2}) }(\cot \theta ), \quad t=n+|{\widetilde m}|.
\label{MPH3}
\end{eqnarray}
The $B_t$ matrices in (\ref{MPH1}) in first instance describe the decomposition of the non-exponential part of 
${\mathcal F}_t^{|{\widetilde m}|}(\theta ) $ into associated Legendre functions, $P_t^{|m|}(\theta)$, with $|m|\in \left[ |{\widetilde m}|, t\right]$.
Explicit expressions are given in \cite{MolPhys}.
Though the $t$-label in the last equation no longer has the meaning of ordinary angular momentum, it nonetheless
continues labeling the $so(3)$ irreps in the new (potential) representation of the algebra, and ${\widetilde m}$,
continues obeying same branching rule with respect to $t$ as does the ordinary  magnetic quantum number $m$ with respect to 
conventional orbital angular momentum, $l$.
However, the Euclidean distance, $ r^2=x^2+y^2+z^2$, is no longer conserved by
transformations generated by the  algebra elements in the new representation. 
Rather, one encounters the deformed metric,
\begin{equation}
(x^2+y^2+z^2)\longrightarrow e^{-2b\theta}(x^2+y^2+z^2).
\label{def_S2}
\end{equation}
The representation functions ${\mathcal X}_t^{\widetilde{m}}(\theta, \varphi)$ of the non-unitarily transformed $ so(3)$
in eqs.~ (\ref{MPH2}), (\ref{MPH3}) are defined in terms of non-classical Romanovski polynomials \cite{raposo},
$R_n^{\alpha, \beta }(\cot \theta)$, which satisfy the following hyper geometric differential equation,
\begin{eqnarray}
(1+x^2)\frac{{\mathrm d}^2R_n^{\alpha, \beta}}{{\mathrm d} x^2}
+2\left(\frac{\alpha }{2} +\beta x
\right)\frac{{\mathrm d}R_n^{\alpha, \beta}}{{\mathrm d}x}
-n(2\beta +n-1)R_n^{\alpha, \beta}=0,
\end{eqnarray}
and are
obtained from the  weight function
\begin{eqnarray}
\omega ^{\alpha, \beta}(x)=(1+x^2)^{\beta -1}\exp(-\alpha \cot^{-1}x),
\end{eqnarray}
by means of the Rodrigues formula:
\begin{eqnarray}
R^{\alpha, \beta}_n(x)=\frac{1}{\omega^{\alpha, \beta }(x)}
\frac{{\mathrm d}^n}{{\mathrm d}x^n}
\left[ (1+x^2)^n\omega^{\alpha, \beta}(x)\right].
\end{eqnarray}
The parameters $\alpha $ and $\beta$  (labeled by the index $t$ in the text) are 
\begin{eqnarray}
\alpha =\frac{2b}{t+\frac{1}{2}}, &&
\beta =-\left(t+\frac{1}{2}\right) +1, \quad t=n+|{\widetilde m}|.
\end{eqnarray}
In ref.~\cite{MolPhys} it has been shown that, due to the presence of the representation constant $\alpha_t$ in the expression for 
the energy in (\ref{MPH1}),  the cotangent perturbed rigid rotator describes rotational bands of anomalously large gap 
between the ground and the first excited sates, as illustrated in Fig.~2.
In recapitulation of the above examples, it must have become clear that 
the potential representation of an algebra  can be given the meaning of metric deformation
of the initial group symmetry space, and thereby of a group symmetry breaking
by the mass scale introduced by the potential strength. In the subsequent section we shall use
the potential algebra concept as a guidance in the search for a potential of a conformal symmetry algebra which we then will 
employ as a quark potential with the aim to explain the observed conformal degeneracies in the unflavored hadron spectra.

\begin{figure}
\includegraphics[height=.3\textheight]{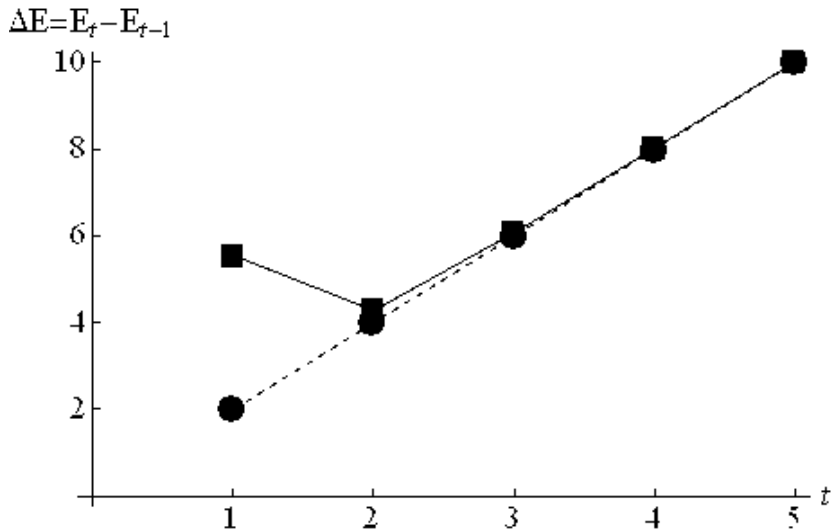}
\caption{The splitting, $\Delta E=(E_{t}-E_{t-1})$,  for a moderate $b=1$ value,
between neighboring rotational levels belonging to
the ordinary rigid rotator (circles connected by a dashed line)
and the cotangent-perturbed rigid rotator (squares connected by a solid line). 
The rotational bands of the perturbed rotator show a typical
anomalous splitting between the ground state and its first excitation.
For the higher lying excitations  the perturbed level splittings 
rapidly approach the unperturbed ones, a behavior visualized by 
practically coincident squares and circles.}
\end{figure}

\section{Conformal symmetry in QCD  and degeneracies in hadron spectra}
The unflavored baryon spectra reported so far reveal typical conformal degeneracies, as visible from the fact
that the respective nucleon $(N$), and $\Delta(1232)$-excitations with masses between 1250 MeV and 2500 MeV distribute
each over three $so(4)$ levels of the type, $(K/2,K/2)\otimes \left[ (1/2,0)\oplus (0,1/2) \right]$
with K=1,3, and 5 \cite{KiMoSmi}. Levels with K=2,4 are ``missing''. Such level sets can be viewed as pieces of
infinite dimensional unitary $so(2,4)$ irreducible representations and are indicative of the relevance of
conformal symmetry in QCD \cite{KiCo}. The QCD is conformal in two regimes, in the ultraviolet (UV) where the partons are
practically  mass-less and the strong coupling tends to zero, and in the infrared (IR) where according to
recent measurements \cite{Doer}, the strong coupling walks toward a fixed value (so called ``conformal window'').
At the mass scale of interest, the $u$ and $d$ quark masses can be viewed as small and one may consider the
possibility that the aforementioned degeneracies represent a print left by the conformal symmetry.
The conformal symmetry of QCD is furthermore supported by the gauge/gravity duality \cite{Maldacena}
which identifies the conformal gauge theory at the boundary of the $AdS_5$ space-time with QCD at high-temperatures.
The $AdS_5$ geometry is $S^1\otimes S^3$ and the isometry algebra  of $S^3$ is $so(4)$, a reason for
which the excitation modes of the free motion on $S^3$  are classified according to the $so(4)$ irreps,
and the energies are $(K+1)^2$-fold degenerate. This is so because free motion on $S^3$  describes 4D rotational bands
in terms  of the eigenvalue problem
of the squared 4D angular momentum, ${\mathcal K}$, according to ${\mathcal K}Y_{Klm}(\chi, \theta, \varphi)=\epsilon_K  Y_{Klm}(\chi, \theta, \varphi)$,
with $\epsilon_K=K(K+2)$, and $Y_{Klm}(\chi,\theta, \varphi)$ standing for the 4D hyper spherical harmonics. 
Here, $\chi$, and $\theta$ denote the two 
polar angles, while $\varphi$ is the azimuthal angle. The $S^3$ metric is given by $ds^2=d\chi^2 +\sin^2\theta (d\theta  +\sin^2\varphi d\varphi )$.
Comparison of these 4D rotational bands with data on the $N$ and $\Delta$ excitations reveals a good match between the quantum numbers, but
the respective bands are characterized by  anomalous gaps between the ground state and the first excited state, as visible from Fig.~3.
In order to describe this observation we  approximate, for the time being, 
the hadrons by two-body systems, $q-(qq)$ for baryons, and $q-\bar q$ for mesons, 
and  replace them afterward by  scalar particles on $S^3$ of the corresponding reduced mass, which we 
treat  as free parameters.
\begin{figure}
 \includegraphics[height=.3\textheight]{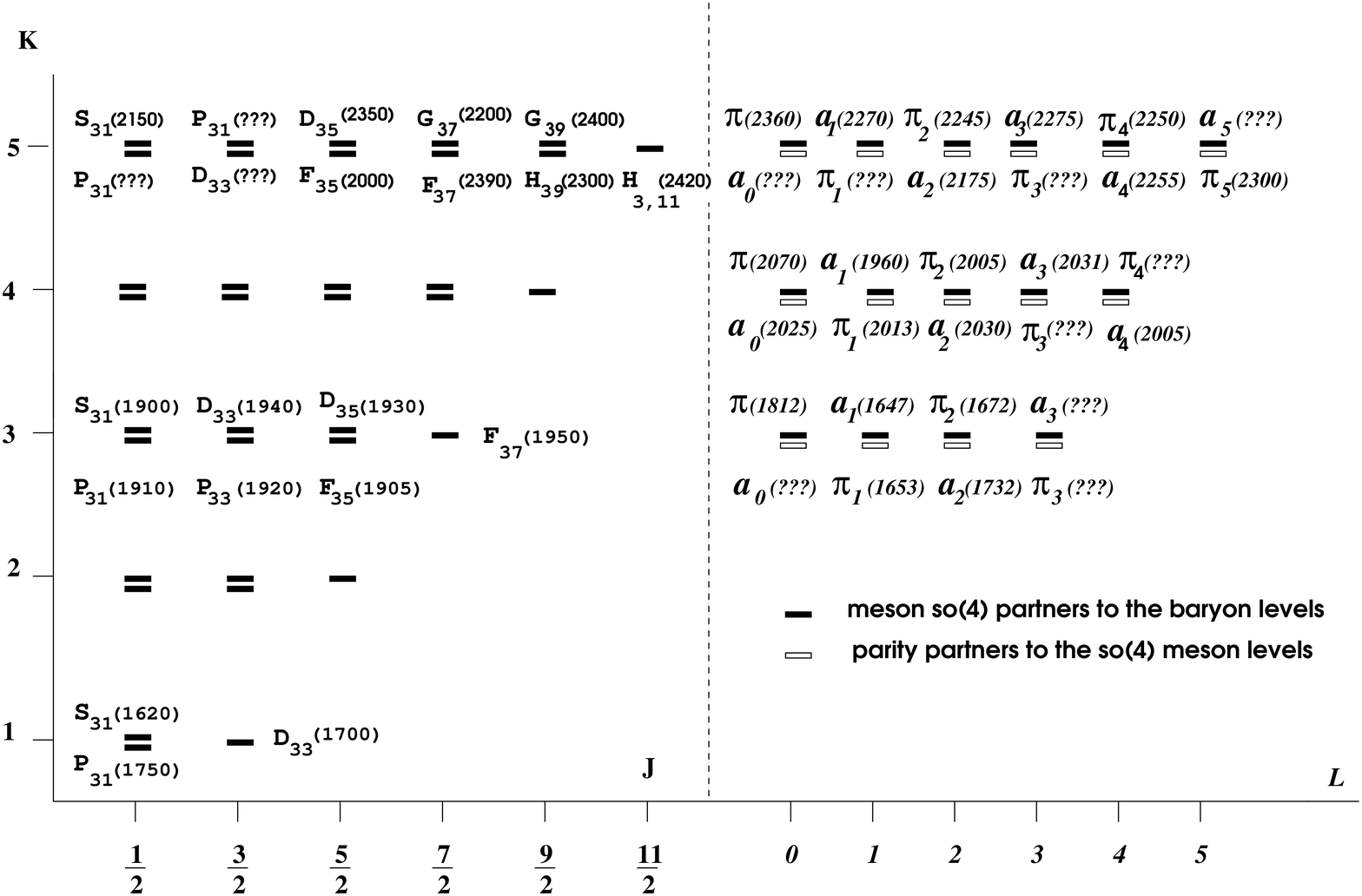} 
\hspace{0.2cm}
\includegraphics[height=.27\textheight]{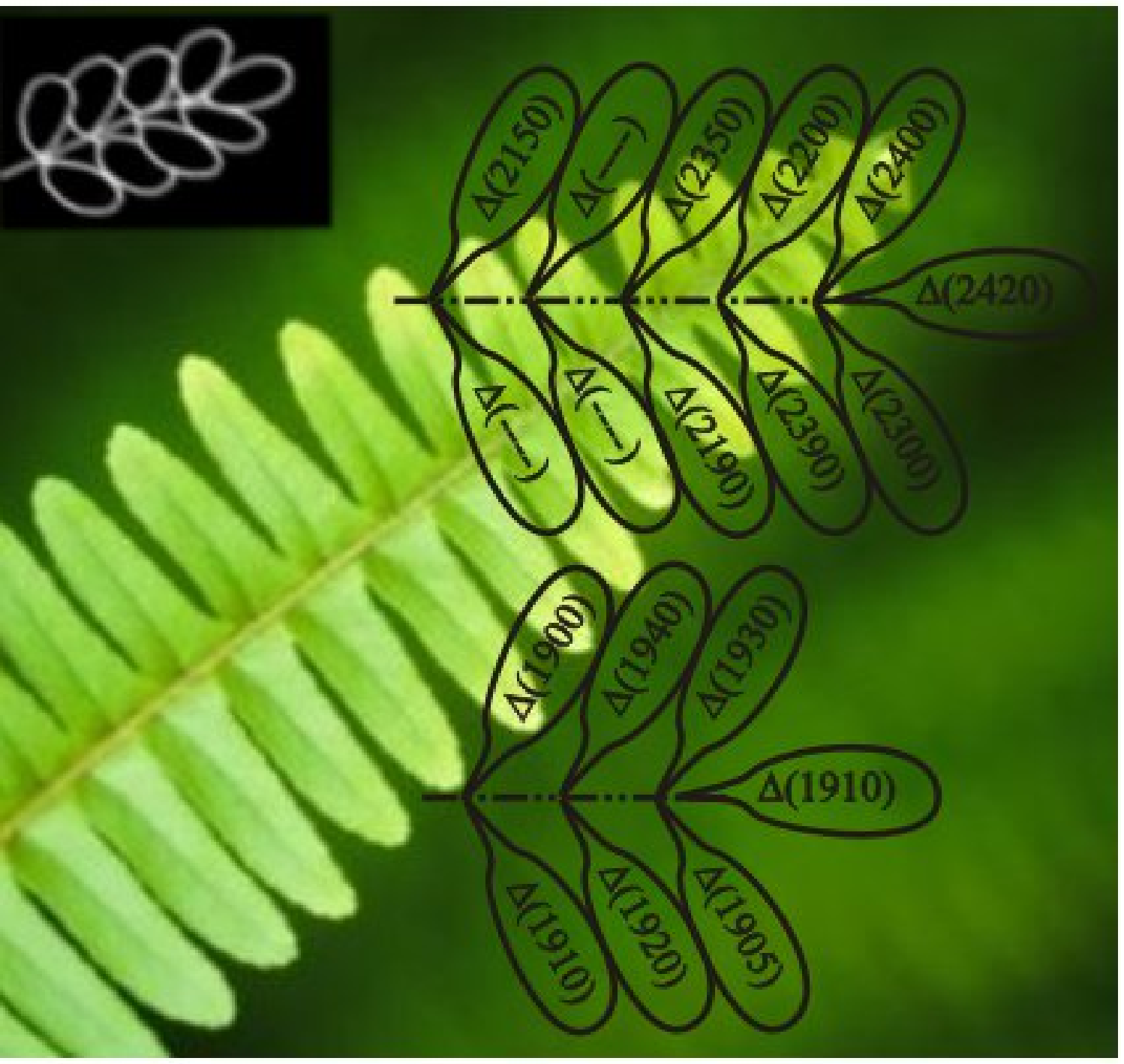}  
  \caption{$\Delta (1232)$ spectrum on the plane spin (J), versus  4D angular momentum, $K$ (replaced by $(K+1)$) together
with excited isovector mesons according to the Crystal Barrel data,  compiled in \cite{Afonin_SS}. 
Each baryon level has $K$ bifurcations into states of opposite parities, and contains a single-parity state of the maximal spin
$J=(K+1/2)$. In resemblance of fern, we choose to term such multi-spin-parity clusters
as spin-(K+1/2) {\it ferneons} (schematically represented in the right figure, courtesy C. Compean).  
The gap of $\sim$500 MeV between $\Delta(1232)$ and the spin- 3/2 ferneon
is significantly larger than the subsequent gaps of $\approx 200$ MeV among ferneons whose spins differ by two units, 
a behavior which hints on the necessity of introducing on $S^3$ a potential of an $so(4)$ symmetry algebra, 
in parallel  to the lower dimensional  $S^2$ case in Fig.~2. The spin-$5/2^+$ and $9/2^+$  ferneons are ``missing''. 
The figure is suggestive of a fermion-boson supersymmetry in the spirit of \cite{FIach}.
The $N$ spectrum follows similar patterns (see  \cite{KiMoSmi} for details).}
\end{figure}
\noindent 
On the basis of the experience made in the previous section it is not difficult to realize
that it is the cotangent-perturbed 4D rigid rotator on $S^3$
that can be expected to generate 4D rotational bands
with the required anomalous gap between the ground state and its first excitation mode.
The $\cot\chi$  potential, first introduced by Schr\"odinger \cite{Schr41} as a harmonic function on $S^3$, derives its utility for employment in the 
description of hadron spectra from its close  relationship to 
the Cornell potential,  predicted by lattice QCD,
as visible from  its Taylor series decomposition \cite{TQC},
\begin{equation}
-2b\cot \frac{\stackrel{\frown}{r}}{R}
= -\frac{2bR}{\stackrel{\frown}{r}}
+\frac{4b}{3R}\stackrel{\frown}{r}+..., \quad \mbox{for}\quad
\chi=\frac{\stackrel{\frown}{r}}{R},
\label{Cornell}
\end{equation}
where $\stackrel{\frown}{r}$ denotes the geodesic distance on $S^3$, and $R$
is the $S^3$ radius. Conformal symmetry implies $R$ independence.  
The $so(4)$ symmetry algebra of this potential has been explicitly constructed in \cite{Adrian} following the scheme
explained in the previous section, namely, establishing the equivalence 
\begin{eqnarray}
\left({\mathcal K} -2b\cot\chi \right)\Psi_{K{\widetilde l}{\widetilde m}}(\chi, \theta, \varphi)=
\left(\widetilde{{\mathcal K}} -\frac{\alpha_K^2}{4}\right)\Psi_{K{\widetilde l}{\widetilde m}}(\chi, \theta, \varphi) &=&
\left( (K(K+2) -\frac{\alpha_K^2}{4}\right)\Psi_{K{\widetilde l}{\widetilde m}}(\chi, \theta, \varphi),\nonumber\\
{\Psi_{K{\widetilde l}{\widetilde m}}(\chi,\theta,\varphi)}=
e^{\frac{\alpha_K\chi}{2}}\psi_K^{\widetilde l}
(\chi)Y_{\widetilde l}^{\widetilde m}(\theta,\varphi), \quad
\psi_K^{\widetilde l}(\chi)=\sin^K\chi R_{K-{\widetilde l}}^{\alpha_K,\beta_K-1}
(\cot\chi) ,& \quad&
\alpha_K=-\frac{2b}{K+1}, \quad \beta_K=-K,
\label{RM_2}
\end{eqnarray}
with
\begin{eqnarray}
\widetilde{{\mathcal K}}&=& e^{\frac{\alpha_K\chi}{2}}{\mathbf A}_K{\mathcal K} e^{-\frac{\alpha_K\chi}{2}}{\mathbf A}_K^{-1}.
\end{eqnarray}
The matrices ${\mathbf A}_K$ realize in first instance  the expansions of the  
$\psi_K^{\widetilde l}(\chi)$ functions into $\sin^l\chi {\mathcal G}_{K-l}^{l+1}(\cos\chi)$, with $l\in \lbrack {\widetilde l},K\rbrack $,
and ${\mathcal G}$ being the Gegenbauer polynomials.
The eigenvalue problem of the Casimir invariant $\widetilde{{\mathcal K}}$ of the non-unitarily transformed $so(4)$ algebra
is equivalent to the $\cot\chi$ perturbed rigid rotator on $S^3$, the symmetry space of the 
orthogonal group $SO(4)$.  The line element on the exponentially rescaled $S^3$ has been found  as \cite{Raya},
\begin{eqnarray}
{\mathrm d}^2 {\widetilde s}&=&
e^{-b\chi}\left( 
(1+b^2/4) {\mathrm d}^2\chi +
\sin^2\chi \left( {\mathrm d}\theta^2 +\sin^2\theta {\mathrm d}^2\varphi
\right)\right).
\label{def_me}
\end{eqnarray}
The latter equation shows that the energy scale introduced by the  potential strength provides the deformation of 
the $SO(4)$ symmetry space. As long as $SO(4)\subset SO(2,4)$, also the conformal group symmetry will be broken by same scale.
On the other side, conformal symmetry is broken by the dilaton mass and it is worth fitting the potential strength to data with the aim
to see how the potential strength compares to the dilaton mass. In \cite{Raya} such an analysis has been performed on the basis
of the data on the high-lying unflavored meson excitations, in the compilation of \cite{Afonin}. We here show in Fig.~4 the data
on the isotriplet vector meson excitations for illustrative purposes. The rest of the data can be consulted in \cite{Raya}.
The data on the spectra of the high-lying unflavored mesons 
show not only pronounced conformal grouping  patterns above 
$\approx$ 1250 MeV but also the fermion-meson degeneracies shown in Fig.~3. 
Our case is that these patterns emerge in consequence of
the breaking of the $SO(2,4)$ group symmetry at the level
of the representation of its algebra as made evident through eq.~(\ref{def_me}),
and conjecture that  such a breaking may be attributed to the dilaton.
In \cite{Raya} a $b$-value of $b=3.2793\pm 0.0697$ has been obtained, which is to a very good approximation
isospin independent, a circumstances that speaks in favor of a possible universality. Also the $S^3$ radius, $R$ has been treated as
a free parameter, with the aim to obtain its inverse, $1/R$, which  defines the finite temperature. In units of mass, one finds,
$\hbar c\sqrt{b}/R =673.7$ MeV,  which hits the size of the light dilaton mass predicted in \cite{MD}. Also the value calculated for 
the temperature, T=$\hbar c/R$=373 MeV, seems reasonable insofar as it is significantly, though not much larger, 
than $\Lambda_{QCD}\approx 175$ MeV.
 \begin{figure}
\includegraphics[height=.3\textheight]{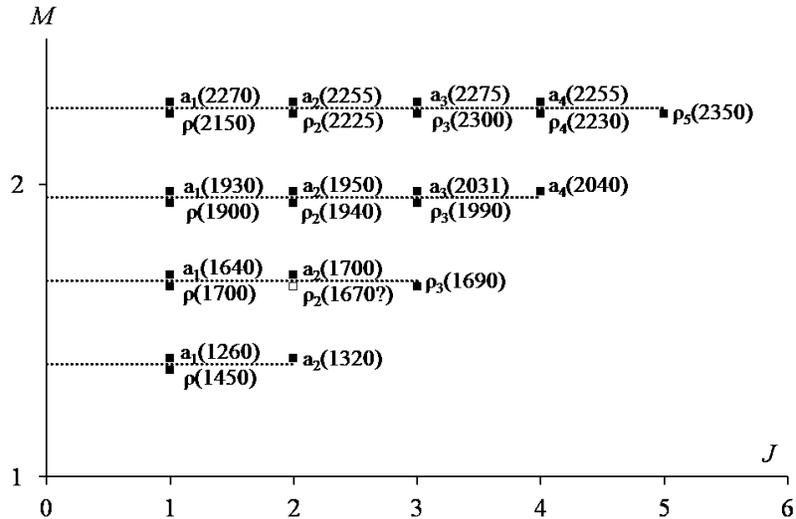}
\caption{High-lying excitations of the isotriplet vector  
mesons according to the compilation of 
\cite{Afonin} which places mesons of opposite $G$ and $C$ parities in same so(4)
level. Remarkable, the narrow parity doubling of the levels, likely to be  indicative of the chiral symmetry restoration from the spontaneously broken Goldstone-- to the manifest Wigner mode in this excitation region. }
\end{figure}
Our conclusion is that the conformal symmetry in the  unflavored sector in QCD may throughout  be broken by the dilaton mass
in the subtle way at the level of the group and conserved as potential algebra of the interaction, thus preserving the conformal degeneracies.
The observed and described conformal grouping  patterns
in the 1250--2500 MeV excitation region of the unflavored hadrons, both baryons \cite{KiCo} and mesons, 
are in our opinion representative at least for the realization of a  partial conformal dynamical symmetry in the spirit of \cite{Levine}.
Our findings are suggestive of a generalization of the traditional algebraic  Hamiltonians \cite{YachellloSo4}
from $H=\sum_r {\mathcal C}^r$ to ${\mathcal H}=\sum_p\left( {\mathbf F}{\mathcal C}{\mathbf F}^{-1}+\gamma_\ell \right)^p$, 
for suited non-unitary ${\mathbf F}$'s, and properly introduced representation constants, $\gamma_\ell$ where $\ell$ is a generic representation
label. Such a scheme could be helpful in improving the description of the
wave functions of a variety of systems by replacing the representation functions of  $SO(n)/SO(p,q)$ groups
by more adequate expressions.

\section*{Acknowledgments}
The efforts of the organizers in making possible the memorable event {\it Beauty in Physics:Theory and Experiment}, May 14-18,  
2012 at Cocoyoc, Mexcio, where this talk was given to honor the 70th birthday of Francesco Yachello and his trailblazing work,
are highly appreciated. Special thanks to David Edwin Alvarez Castillo, Cliffor Benjamin Compean Jasso, Nehemias Leija Martinez, Adrian Pallares Rivera,
Alfredo Raya, Alvaro Per\'ez Raposo, and Hans J\"urgen Weber for invaluable collaboration and support over the last eight years.
I am furthermore indebted to Jose Limon Castillo for permanent assistance in computer matters.

\end{document}